\PassOptionsToPackage{comma,numbers,sort&compress,super}{natbib}  

\documentclass[%
 onecolumn,  
superscriptaddress,
preprint,
nofootinbib,
 aps,
]{revtex4-2}

\usepackage{import}
\usepackage{usepackets/usepackets}
\usepackage[normalem]{ulem}
\usepackage{array, makecell} 
\setcitestyle{super}

\usepackage[articletitle=true, etalmode = truncate, maxauthors = 10]{achemso}

\newcommand{\rev}[1]{{\color{black}{#1}}}

\begin{document}

\title{Quantum  Equation of Motion with Orbital Optimization for Computing Molecular Properties in Near-Term Quantum Computing }

\author{Phillip W. K. Jensen}
 \email{pwkj@chem.ku.dk}
\affiliation{Department of Chemistry, University of Copenhagen, DK-2100 Copenhagen \O.}

\author{Erik Rosendahl Kjellgren}
\affiliation{Department of Physics, Chemistry and Pharmacy,
University of Southern Denmark, Campusvej 55, 5230 Odense, Denmark. }

\author{Peter Reinholdt}
\affiliation{Department of Physics, Chemistry and Pharmacy,
University of Southern Denmark, Campusvej 55, 5230 Odense, Denmark. }

\author{Karl Michael Ziems}
\affiliation{Department of Chemistry, Technical University of Denmark, Kemitorvet Building 207, DK-2800 Kongens Lyngby, Denmark.}

\author{Sonia Coriani}
\affiliation{Department of Chemistry, Technical University of Denmark, Kemitorvet Building 207, DK-2800 Kongens Lyngby, Denmark.}

\author{Jacob Kongsted}
\affiliation{Department of Physics, Chemistry and Pharmacy,
University of Southern Denmark, Campusvej 55, 5230 Odense, Denmark. }

\author{Stephan P. A. Sauer}
 \email{sauer@chem.ku.dk}
\affiliation{Department of Chemistry, University of Copenhagen, DK-2100 Copenhagen \O.}

\date{\today}

\begin{abstract}
Determining the properties of molecules and materials is one of the premier applications of quantum computing. A major question in the field is how to use imperfect near-term quantum computers to solve problems of practical value. Inspired by the recently developed variants of the quantum counterpart of the equation-of-motion (qEOM) approach and the orbital-optimized variational quantum eigensolver (oo-VQE), we present a quantum algorithm (oo-VQE-qEOM) for the calculation of molecular properties by computing expectation values on a quantum computer. We perform noise-free quantum simulations of BeH$_2$ in the series of \mbox{STO-3G/6-31G/6-31G*} basis sets and of H$_4$ and H$_2$O in 6-31G using an active space of four electrons and four spatial orbitals (8 qubits) to evaluate excitation energies, electronic absorption, and, for twisted H$_4$, circular dichroism spectra. We demonstrate that the proposed algorithm can reproduce the results of conventional classical CASSCF calculations for these molecular systems.

\end{abstract}

\maketitle


\section{Introduction}

Powerful quantum algorithms have been developed for solving problems in quantum chemistry and 
material science for future fault-tolerant quantum computers. These problems include estimating ground-~\cite{efficient_lee_2021, su_fault-tolerant_2021,joshua_reliably_2022} and excited-state energies~\cite{bauman_toward_2021,jensen_quantum_2021,raffaele_witnessing_2018}, molecular gradients~\cite{OBrian_efficient_2022},  response properties of molecules~\cite{cai_quantum_2020} and many others, with the potential to realize a runtime advantage compared to the classical counterparts~\cite{joshua_reliably_2022,efficient_lee_2021, babbush_focus_2021}. Indeed, recent progress has demonstrated the ability of controlled quantum systems to perform tasks beyond the capabilities of even large supercomputers~\cite{arute_quantum_2019,zhong_quantum_2020,wu_strong_2021}. However, many challenges remain in making quantum computers capable of running advanced quantum algorithms with direct applicability. Current quantum computing hardware is still limited in size and tends to be prone to noise that corrupts information and prohibits long run-times~\cite{bharti_noisy_2022}. While noise can be managed using quantum error correction~\cite{gottesman_introduction_2009}, doing so will require redundant encoding of the information, thus further restricting the effective size of implementable algorithms. This leads to the question of whether efficient and reliable property estimation is feasible on near-term quantum computers. With recent advances in quantum hardware~\cite{kim_evidence_2023}, researchers have been investigating methods for using near-term quantum computers to obtain molecular properties~\cite{huang_variational_2022, ollitrault_estimation_2023}. For the area of quantum chemistry and materials, much of this work has focused on the task of ground state energy estimation, with the primary method being the variational quantum eigensolver (VQE) algorithm~\cite{mcclean_theory_2016,peruzzo_variational_2014}. Yet, there is widespread interest in calculating the energy spectrum of a Hamiltonian, for example, to understand optical spectra in quantum chemistry. To address this, the original VQE scheme was extended in the following two ways. First, it was combined with orthogonality constraints \cite{higgott_variational_2019, jones_variational_2019, ryabinkin_constrained_2018,yen_exact_2019} to enable systematic exploration of the low-lying excited states. Second, quantum equation-of-motion (qEOM) techniques were proposed for exploration of the low- and the high-lying excited state and energies~\cite{mcclean_hybrid_2017, ollitrault_quantum_2020}. \rev{Since then various variants and applications of the qEOM method have emerged. These include the computation of one-particle Green's functions~\cite{rizzo_one-particle_2022}, multicomponent qEOM variants~\cite{pavošević_multicomponent_2021, pavošević_polaritonic_2021}, molecular properties~\cite{nakagawa_analytical_2023,asthana_quantum_2023, kumar_quantum_2023,kim_two_2023}. An overview of qEOM-based methods is presented in Table \ref{tbl:overview}}. Other techniques, such as polynomial expansion, have also been proposed to approximate spectral functions and extract excited state energies~\cite{jamet_krylov_2021,jensen_near-term_2023}. However, all these methods rely on the standard VQE observable estimation routines which is likely to suffer from a measurement bottleneck~\cite{gonthier_measurements_2022}. The measurement problem in VQE can be mitigated by alleviating the number of measurements, e.g., using informationally complete positive operator-valued measurements~\cite{garcia_learning_2021, glos_adaptive_2022}, grouping techniques~\cite{yen_measureing_2020, yen_catan_2021,choi_improving_2022}, or shadow tomography~\cite{huang_predicting_2020}.


 In classical computing, many benchmark studies have been carried out for both highly accurate and economical methods for the calculation of excitation energies of small and medium size organic molecules starting, e.g., from the Thiel benchmark set~\cite{schreiber_caspt2_2008, sauer_cc3_2009, silva-junior_cc3_2010, silva-junior_cc3_2010, sauer_soppa_2015} with CASPT2, CC2, CCSD and CC3 calculations to the QUEST database~\cite{loos_quest_2020,veril_quest_2021}, which nowadays even includes CCSDT, CCSDTQ and CC4 results~\cite{loos_cc4_2021,loos_cc4_2022}. Although accurate classical quantum chemical methods, such as the ones based on coupled-cluster theory, offer a very appealing hierarchy of methods due to their size-extensivity and systematic hierarchy of improvability~\cite{helgaker_molecular_2000,helgaker_response_2012}, the most accurate methods are also associated with steep computational costs, which may render them unusable beyond relatively small molecular systems. In quantum computing, coupled-cluster calculations can be realized using its unitary version of the exponential ansatz, i.e., unitary coupled cluster (UCC) theory~\cite{jarrod_theory_2016, strategies_romero_2019, anand_quantum_2022}. UCC exhibits important features such as unitarity and a variational character, and it is believed that UCC results in a more multi-configurational wave function compared to coupled-cluster~\cite{helgaker_molecular_2000, christiansen_coupled_2006, bartlett_coupled-cluster_2012}. This is important because several interesting chemical problems pertain to large and complicated multiconfigurational systems, especially within life sciences. UCC does however suffer from relatively deep circuits. In classical computing, the complete-active space self-consistent field (CASSCF)~\cite{siegbahn_comparision_1980,roos_complete_1980,siegbahn_complete_1981}, complete-active space configuration interaction (CASCI)~\cite{olsen_determinant_1988}, and density matrix renormalization group (DMRG)~\cite{white_density_1992,white_density-matrix_1993,baiardi_dmrg_2020} are all used to treat molecular systems with multiconfigurational character. The quantum counterpart to CASSCF is the orbital-optimized VQE~\cite{bierman_improving_2023, sokolov_quantum_2020,mizukami_orbital_2020} (oo-VQE) method, where oo-VQE displays a potential advantage over CASSCF by using a quantum device to compute the energy, which on future QPUs could enable the method to treat larger active spaces where CASSCF struggles due to its exponential resource demands.


Our work explores how near-term quantum computers can be used to simulate the spectroscopic properties of molecules. To achieve this, we propose a quantum equation-of-motion (qEOM) based algorithm for the calculation of molecular properties, specifically excitation energies, electronic absorption and circular dichroism spectra, following the setup presented by Ollitrault \emph{et al.}~\cite{ollitrault_quantum_2020}, where the ground state of the system is prepared via the VQE and the matrix elements of the EOM are then estimated directly on a quantum computer. The resulting matrix is diagonalized on a classical computer to extract the excitation energies and the excited states. Our method is similar in that we also explicitly construct the EOM matrix and diagonalize it on a classical computer, but we use the theoretical framework of oo-VQE-qEOM with an active space. Using an active space reduces the number of qubits and gates needed for the simulation, which makes our method more suitable for NISQ devices. The oo-VQE approach has so far only been formulated for ground-state properties such as polarizabilities~\cite{nakagawa_analytical_2023}. In this work, we combine the oo-VQE algorithm with qEOM to obtain properties relating directly to excited states.

\section{Theory}
In the following, we outline the theoretical background for our developments. First, we give an overview of the VQE algorithm in the active space approximation; second, the orbital-optimized VQE is briefly discussed; third, we combine oo-VQE and qEOM to calculate molecular properties.  Finally, we briefly discuss why the recently proposed quantum self-consistent equation-of-motion method~\cite{asthana_quantum_2023} is unsuitable for near-term hardware using our setup.

\subsection{The active space approximation}

In the active-space approximation, the wave function can be written as 
\begin{equation}
    \left|\Psi(\vec{\theta})\right> = \left|I\right>\otimes \left|A(\vec{\theta})\right>\otimes \left|V\right>, \label{eq:wf_tot}
\end{equation}
where $\ket{I}$ denotes the inactive orbitals and can be expanded as $ \ket{I} = \ket{1}\otimes\ket{1} \otimes ... \otimes \ket{1}$, $\ket{V}$ denotes the virtual orbitals, $ \ket{V} = \ket{0}\otimes  \ket{0} \otimes ... \otimes \ket{0}$, and the active part, $  \ket{A(\vec{\theta})} $, is given as
\begin{equation}
    \left|A(\vec{\theta})\right> = U(\vec{\theta})\left(\left|1\right>\otimes \left|1\right> \otimes ...\otimes \left|0\right> \otimes \left|0\right>\right),
\end{equation}
where $\vec{\theta}$ are the quantum circuit parameters, such as qubit rotation angles. The advantage of using the active space approximation is that we can remove the qubits corresponding to the inactive and virtual parts, which leaves only the active space to be simulated on a quantum computer.

In this work, we will consider the molecular (non-relativistic) singlet electronic Hamiltonian, which in second quantized form is given by~\cite{helgaker_molecular_2000}

\begin{align}
\hat{H}\left(\vec{\kappa} \right) = \sum^N_{pq} h_{pq}\left(\vec{\kappa} \right) \hat{E}_{pq}  + \frac{1}{2} \sum^N_{pqrs} g_{pqrs}\left(\vec{\kappa} \right) \left( \hat{E}_{pq} \hat{E}_{rs} - \delta_{qr}\hat{E}_{ps}  \right) + h_{\text{nuc}}\label{eq:fermionic_hamil}
\end{align}
where  $\hat{E}_{pq} = \hat{a}^\dagger_{p\alpha}\hat{a}_{q\alpha} + \hat{a}^\dagger_{p\beta}\hat{a}_{q\beta}$ is the usual (spin-adapted singlet) one-electron excitation operator where $\alpha$ and $\beta$ indicate the orientation of spin,  \emph{N} is the number of spin-orbitals, $h_{pq}$ and $g_{pqrs}$ are the one- and two-electron integrals, $\vec{\kappa}$ are the orbital rotation parameters that determine the molecular orbitals, and $h_{\text{nuc}}$ is the nuclear repulsion energy. The Hamiltonian can be expressed as a linear combination of unitaries 
\begin{align}
\hat{H}\left(\vec{\kappa} \right) = \sum^L_i c_i(\vec{\kappa}) \hat{P}_i~, \label{eq:hamil}
\end{align}
where $L = O(N^4)$ and $\hat{P}_i$ are strings of Pauli operators. The coefficients $c_i(\vec{\kappa})$ depend on the one- and two-electron integrals used to generate the fermionic Hamiltonian \eqref{eq:fermionic_hamil} and the coefficients from the mapping of fermionic to Pauli operators. The Pauli strings can be partitioned as:
\begin{align}
 \hat{P}_i = \hat{P}^{(i)}_I \otimes \hat{P}^{(i)}_A \otimes \hat{P}^{(i)}_V, \label{eq:expansion_pauli}
\end{align}
where $\hat{P}^{(i)}_I$ acts on the qubits in the inactive space,  $\hat{P}^{(i)}_A$  on the active space, and $\hat{P}^{(i)}_V$ on the virtual space. The energy can then be written as 
\begin{align}
E\left(\vec{\kappa}, \vec{\theta}\right) &=    \left<\Psi(\vec{\theta})\right| \hat{H}\left(\vec{\kappa} \right)\left|\Psi(\vec{\theta})\right>\\
&= \sum^L_i c_i\left(\vec{\kappa}\right) \left<I\left|\hat{P}^{(i)}_{I}\right|I\right>
\left<A(\vec{\theta})\left|\hat{P}^{(i)}_{A}\right|A(\vec{\theta})\right> \left<V\left|\hat{P}^{(i)}_{V}\right|V\right>. \label{eq:energy}
\end{align}
If $\hat{P}^{(i)}_{I}$ or $\hat{P}^{(i)}_{V}$ contain the Pauli operators $\hat{\sigma}_X$  or $\hat{\sigma}_Y$,  the corresponding term in 
Eq. \eqref{eq:energy} vanishes and can be removed from the equation. Next, we remove the virtual part because for the remaining terms $\left<V\left|\hat{P}^{(i)}_{V}\right|V\right> = 1$ $\forall i$, since $\hat{\sigma}_Z \ket{0} = \ket{0}$. For the inactive space, a sign flip occurs if there is an odd number of Pauli-\emph{Z} operators in $\hat{P}^{(i)}_{I}$, since $\hat{\sigma}_Z \ket{1} = -\ket{1}$. For an even number of Pauli-\emph{Z} operators, nothing happens. The energy expression \eqref{eq:energy} can therefore be written as
\begin{align}
E\left(\vec{\kappa}, \vec{\theta}\right) &= \sum^{L'}_i c_i\left(\vec{\kappa}\right) (-1)^{ \left(\# \hat{\sigma}_Z\right)^{(i)}_I}
\left<A(\vec{\theta})\left|\hat{P}^{(i)}_{A}\right|A(\vec{\theta})\right> \\
&=\sum^{L'}_i \mathcal{C}_i\left(\vec{\kappa}\right) 
\left<A(\vec{\theta})\left|\hat{P}^{(i)}_{A}\right|A(\vec{\theta})\right>\label{eq:energy_1},
\end{align}
where $\mathcal{C}_i\left(\vec{\kappa}\right) = c_i\left(\vec{\kappa}\right) (-1)^{ \left(\# \hat{\sigma}_Z\right)^{(i)}_I}$, and $ \left(\# \hat{\sigma}_Z\right)^{(i)}_I$ is the number of Pauli-\emph{Z} operators in the inactive space. The number of qubits in VQE is reduced from \emph{N} to $N_A$, where $N_A$ is the number of spin-orbitals in the active space. Further, the number of terms in the active space Hamiltonian \eqref{eq:energy_1} can be significantly less than the full space Hamiltonian \eqref{eq:hamil}, i.e., $L\gg L'$, resulting in a smaller variance for the energy estimation using the active space Hamiltonian.

\subsection{oo-VQE algorithm} 
\label{sec:oo-VQE}
We will now briefly explain the orbital-optimized VQE (oo-VQE) algorithm, motivated by similar schemes in 
Refs.~\cite{bierman_improving_2023, sokolov_quantum_2020,mizukami_orbital_2020}. The oo-VQE algorithm is the quantum counterpart of the multi-configurational self-consistent field (MCSCF) approach~\cite{helgaker_molecular_2000}. Similar to how the classical EOM approach starts with an MCSCF procedure, the qEOM begins after the oo-VQE step~\cite{szalay_multiconfiguration_2012}. In this section, we explain how to prepare the oo-VQE ansatz.

In oo-VQE, the minimization problem is both over the space of ansatz parameters for the active space and the space of orbital rotations:
%
\begin{align}
E\left(\vec{\kappa}_{\text{opt}}, \vec{\theta}_{\text{opt}}\right) &= \min_{\vec{\theta}, \vec{\kappa}}  ~  \left<0_{\text{ref}}\left|\mathrm{e}^{-\hat{\sigma}(\vec{\theta})}\mathrm{e}^{-\hat{\kappa}(\vec{\kappa})} \hat{H}\mathrm{e}^{\hat{\kappa}(\vec{\kappa})}\mathrm{e}^{\hat{\sigma}(\vec{\theta})} \right |0_{\text{ref}}\right>  \label{eq:min_energy_1} \\
&= \min_{\vec{\theta}, \vec{\kappa}}\sum^{L'}_i \mathcal{C}_i\left(\vec{\kappa}\right) 
\left<A(\vec{\theta})\left|\hat{P}^{(i)}_{A}\right|A(\vec{\theta})\right> \label{eq:min_energy}
\end{align}
where $\vec{\kappa}_{\text{opt}}$ and $ \vec{\theta}_{\text{opt}}$ denote the optimal parameters, $\ket{0_{\text{ref}}}$ denotes the reference state, e.g., the Hartree-Fock state, $\hat{\sigma}$ is an anti-Hermitian generator, e.g., the Unitary Coupled-Cluster $\hat{\sigma} = \hat{T}-\hat{T}^\dagger$~\cite{strategies_romero_2019,anand_quantum_2022}, which acts within the active space, and $\hat{\kappa}$ is the generator of the orbital rotation operator. The latter is defined as 
\begin{align}
\hat{\kappa} = \sum_{pq} \kappa_{pq} \left( \hat{E}_{pq} - \hat{E}_{qp} \right), \quad \hat{E}_{pq} \in \left\{ \hat{E}_{vi},\quad \hat{E}_{ai},\quad \hat{E}_{av} \right\}, \label{eq:kappa_op}
\end{align}
where the index \emph{i} is used for the occupied (inactive) orbitals, the index \emph{v} is used for the active orbitals, and the index \emph{a} is used for the unoccupied (virtual) orbitals. Note that in conventional VQE, the optimization is only over the space of ansatz parameters, $\vec{\theta}$~\cite{jarrod_theory_2016}. 

The effect of the orbital rotation operator, $\mathrm{e}^{\hat{\kappa}}$, is a rotation of the molecular orbitals, which modifies the one- and two-electron integrals in the molecular electronic Hamiltonian \eqref{eq:fermionic_hamil} according to
\begin{align}
\tilde{h}_{p'q'} &=  \sum_{pq}  \left[\mathrm{e}^{\hat{\kappa}} \right]_{p'p}h_{pq}  \left[\mathrm{e}^{-\hat{\kappa}} \right]_{qq'} \label{eq:h1_tilde} \\
\tilde{g}_{p'q'r's'} &=  \sum_{pqrs}  \left[\mathrm{e}^{\hat{\kappa}} \right]_{p'p}  \left[\mathrm{e}^{\hat{\kappa}} \right]_{q'q}g_{pqrs}\left[\mathrm{e}^{-\hat{\kappa}} \right]_{rr'}  \left[\mathrm{e}^{-\hat{\kappa}} \right]_{ss'}. \label{eq:h2_tilde}
\end{align}
The minimization problem in Eq. \eqref{eq:min_energy} is divided into two subproblems: minimizing
the energy with respect to the ansatz parameters, $\vec{\theta}$, and minimizing the energy with respect to orbital rotations parameters, $\vec{\kappa}$. Different strategies can be used as long as both parameter sets are optimized. For example, one strategy is to keep either $\vec{\theta}$ or  $\vec{\kappa}$ fixed while minimizing
the energy with respect to the other. Another strategy is to optimize both parameter sets, $\{\vec{\theta}\}$ and $\{\vec{\kappa}\}$, at the same time. We refer to the computational details in section \ref{sec:com_details} for more information about the settings used in this work. 

\begin{table}
\centering\renewcommand\cellalign{lc}
\setcellgapes{3pt}\makegapedcells
  \caption{Overview of equation-of-motion-based quantum algorithms. Algorithms: quantum equation-of-motion (qEOM), multicomponent equation-of-motion (mcEOM), quantum electrodynamics equation-of-motion (QED-EOM), quantum self-consistent equation-of-motion (q-sc-EOM), quantum linear response self-consistent qLR(sc) and projected qLR(proj), and qEOM spin-flip unitary coupled cluster (qEOM-SF-UCC).  Properties: electronic excitation (EE), ionization potentials (IP), electron affinities (EA), polarizability ($\alpha$), oscillator strengths (OS), rotational strengths (RS), optical rotation (OR), potential energy surface (PES).  }
  \label{tbl:overview}
  \begin{tabular*}{\textwidth}{@{\extracolsep{\fill}}lllllll}
    \hline
      Year & Reference & Algorithm & Properties & Systems studied\\
    \hline
   2020 & Ollitrault \emph{et al.}~\cite{ollitrault_quantum_2020} & qEOM & EE &  $\text{H}_2$, LiH, H$_2$O (STO-3G)  \\
\rev{2021} & \rev{Pavošević \emph{et al.}~\cite{pavošević_multicomponent_2021}} & \rev{mcEOM} & \rev{EE} & \makecell{\rev{$\text{H}_2$ (STO-3G, 6-31G)} \\\rev{PsH (6-31G, 6-311G)}}   \\
\rev{2021} & \rev{Pavošević \emph{et al.}~\cite{pavošević_polaritonic_2021}} & \rev{QED-EOM} & \rev{EE, IP, EA, OS} & \makecell{\rev{$\text{H}_4$ (STO-3G)} }   \\
   2023 & Asthana \emph{et al.}~\cite{asthana_quantum_2023} & q-sc-EOM &  EE, IP, EA & $\text{H}_2$, $\text{H}_4$, LiH, H$_2$O (STO-3G) \\
    2023 & Kumar \emph{et al.}~\cite{kumar_quantum_2023} & \makecell{qLR(sc) \\qLR(proj)}  &\makecell{ EE, OS, RS, OR, $\alpha$ } & $\text{H}_2$, $\text{H}_4$ LiH,  H$_2$O (STO-3G)\\
    2023 & Kim \emph{et al.}~\cite{kim_two_2023} & q-sc-EOM/Davidson    & EE & \makecell{$\text{H}_4$ LiH,  H$_2$O (STO-3G) \\ $\text{H}_2$ (6-31G)}  \\
    2023 & Pavošević \emph{et al.}~\cite{pavošević_spin-flip_2023} & qEOM-SF-UCC    & PES & \makecell{$\text{C}_2\text{H}_4$ (cc-pVDZ) \\Cyclobutadiene (cc-pVDZ)}   \\
    2023 & This work & oo-VQE-qEOM & EE, OS, RS& \makecell{BeH$_2$ (STO-3G, 6-31G, 6-31G*) \\ H$_4$, H$_2$O (6-31G)}   \\
    \hline
  \end{tabular*}
\end{table}


\subsection{qEOM method in active space approximation}

The equation-of-motion (EOM) approach, first derived by Rowe~\cite{rowe_equations-of-motion_1968}, was extensively reviewed~\cite{schaefer_equations_1977, mcweeny_methods_1992} and implemented in a series of electronic structure packages. In the EOM approach, excited states are generated by applying an excitation operator to the ground state. These operators can formally be written as $\hat{O}^\dagger_k = \ket{\Psi_k}\bra{\Psi_0}$, where $\ket{\Psi_0}$ is the ground state of the system and $\ket{\Psi_k}$ is the \emph{k}th excited state of the system. Similarly, a deexcitation operator can be written as $\hat{O}_k = \ket{\Psi_0}\bra{\Psi_k}$, wherefrom it follows that $\hat{O}_k\ket{\Psi_0} =\ket{\Psi_0}\braket{\Psi_k|\Psi_0} = 0$ $\forall k$. The latter is referred to as the \textit{killer condition}~\cite{mcweeny_methods_1992,szekeres_killer_2001}. 

Having access to the exact ground state $\ket{\Psi_0}$ and the exact excitation operator $\hat{O}^\dagger_k$, the \emph{k}th electronic excitation energy can be computed as
\begin{align}
E_{0k}  & =   \frac{\left<\Psi_0 \left| \left[\hat{O}_k, \left[\hat{H},\hat{O}_k^\dagger \right] \right] \right|\Psi_0\right>}{\left<\Psi_0\left|\left[\hat{O}_k, \hat{O}_k^\dagger\right] \right|\Psi_0\right> } =  \frac{\left<\Psi_0 \left| \left[\hat{O}_k, \hat{H},\hat{O}_k^\dagger \right] \right|\Psi_0\right>}{\left<\Psi_0\left|\left[\hat{O}_k, \hat{O}_k^\dagger\right] \right|\Psi_0\right> }
\label{eq:E0k}
\end{align}
where $E_{0k} = (E_k - E_0)$, and $E_0$ is the ground state energy. The symmetrized double commutator is defined as 
\begin{align}
 \left[\hat{O}_k, \hat{H},\hat{O}^\dagger_k \right] = \frac{1}{2} \left( \left[\hat{O}_k, \left[\hat{H},\hat{O}^\dagger_k \right] \right] + \left[\hat{O}^\dagger_k, \left[\hat{H},\hat{O}_k \right] \right] \right),   \label{eq:symmetrized_dou}
\end{align}
and how to go from the left-hand side of Eq. \eqref{eq:E0k} to expressing the energy in terms of the symmetrized double commutator is explained in Section 2 in Ref. \cite{rowe_equations-of-motion_1968}. In the active space approximation, $\ket{\Psi(\vec{\theta}_{\text{opt}})}$ is an approximation to the exact ground state. It is, therefore, convenient to use the symmetrized double commutator in Eq.~\eqref{eq:E0k} since it will result in a Hermitian Hessian. We approximate the excitation operators by linearly expanding them in a basis of elementary excitations. In this work, we are interested in the electronic excitation energies. Therefore, we use number-conserving spin-adapted one- and two-electron excitation and de-excitation operators in the active space and orbital rotation operators outside the active space with coefficients $\{\mathcal{A},\mathcal{B},\mathcal{X},\mathcal{Y}\}$

\begin{align}
\hat{\mathcal{O}}^\dagger_k = \sum_I \left( \mathcal{A}_I^{(k)} \hat{G}_I - \mathcal{B}_I^{(k)} \hat{G}_I^{\dagger} \right) + \sum_I \left( \mathcal{X}_I^{(k)} \hat{q}_I - \mathcal{Y}_I^{(k)} \hat{q}_I^{\dagger} \right), \label{eq_exci_op}
\end{align}
where $\hat{G}_I$ acts in the active space, and $\hat{q}_I$ acts between the spaces. We need to consider both excitation and de-excitation operators because $\ket{\Psi(\vec{\theta}_{\text{opt}})}$ comprises multiple determinants. The $\hat{G}_I$ operators are the singlet single and singlet double excitation operators given as~\cite{paldus_application_1977, piecuch_orthogonally_1989}

\begin{align}
\hat{G}_I \in &\left\{\frac{1}{\sqrt{2}} \hat{E}_{a i},\quad\frac{1}{2} 
 \frac{1}{\sqrt{\left(1+\delta_{ab}\right)\left(1+\delta_{ij}\right)}}\left(\hat{E}_{ai}\hat{E}_{bj} + \hat{E}_{aj}\hat{E}_{bi}\right), \quad \frac{1}{2\sqrt{3}}  \left(\hat{E}_{ai}\hat{E}_{bj} - \hat{E}_{aj}\hat{E}_{bi}\right) \right\}, \label{eq:G_I}
\end{align}
where $a \geq b$ and  $i \geq j$, and the indices \emph{i} and \emph{j} are used for the occupied reference orbitals and the indices \emph{a} and \emph{b} for the unoccupied reference orbitals in the active space. For the two-electron operators, the operator with the `$+$' sign is couple via intermediate singlet spin states, whereas the operator with the `$-$' sign is couple via intermediate triplet spin states. The orbital rotation operators $\hat{q}_I$ are the singlet one-electron excitation operators given as

\begin{align}
\hat{q}_I \in \left\{\frac{1}{{\sqrt{2}}} \hat{E}_{vi},\quad \frac{1}{{\sqrt{2}}} \hat{E}_{ai},\quad \frac{1}{{\sqrt{2}}} \hat{E}_{av}\right\}.  \label{eq:q_I}
\end{align}
Note that the excitation operator, as defined in Eq.~\eqref{eq_exci_op}, is incomplete, having omitted electron excitation operators higher than double in the active space. In this work, the excitation energy is approximated as

\begin{align}
E_{0k}  & \approx   \frac{\left<\Psi\left(\vec{\theta}_{\text{opt}}\right) \left| \left[\hat{\mathcal{O}}_k, \hat{H}\left(\vec{\kappa}_{\text{opt}}\right),\hat{\mathcal{O}}^\dagger_k \right]\right|\Psi\left(\vec{\theta}_{\text{opt}}\right) \right>}{\left<\Psi\left(\vec{\theta}_{\text{opt}}\right)\left|\left[\hat{\mathcal{O}}_k, \hat{\mathcal{O}}^\dagger_k\right] \right|\Psi\left(\vec{\theta}_{\text{opt}}\right)\right> }, 
\label{eq:E0k_SDC}
\end{align}
where the wave function $\ket{\Psi(\vec{\theta}_{\text{opt}})}$ is obtained  from the oo-VQE algorithm (section \ref{sec:oo-VQE}). To summarize, using Eq. \eqref{eq:E0k_SDC},  there are two primary sources of error which translate into error in the computed excitation energies: the state $\ket{\Psi(\vec{\theta}_{\text{opt}})}$ being an approximation to the exact ground state (Appendix \ref{sec:app:oovqe}), and the excitation operator not being complete.

Applying the variational principle to Eq. \eqref{eq:E0k_SDC}, $\delta(E_{0k} ) = 0$, with respect to the amplitudes for the excitation operator Eq. \eqref{eq_exci_op}, we obtain the generalized eigenvalue equation

\begin{align}
 \boldsymbol{E}^{[2]} \vec{v}_k =  E_{0k} \boldsymbol{S}^{[2]} \vec{v}_k,\label{eq:secular_eq}
\end{align}
where

\begin{align}
 \boldsymbol{E}^{[2]} = \begin{pmatrix}
    \boldsymbol{A} & \boldsymbol{B}          \\[0.3em]
      \boldsymbol{B}^* & \boldsymbol{A}^*           
     \end{pmatrix}, \quad \textbf{S}^{[2]} = \begin{pmatrix}
    \boldsymbol{\Sigma} & \boldsymbol{\Delta}          \\[0.3em]
     -\boldsymbol{\Delta} ^* &  -\boldsymbol{\Sigma}^*           
     \end{pmatrix}, \quad  \vec{v}_k = \begin{pmatrix}
     \vec{Z}_k          \\[0.3em]
      \vec{Y}_k    
     \end{pmatrix}. \label{eq:ESg}
\end{align}
The matrices $\boldsymbol{A}$ and $\boldsymbol{B}$ are defined as

\begin{align}
 \boldsymbol{A} = \begin{pmatrix}
    \boldsymbol{A}^{q,q} &\boldsymbol{A}^{q,G}         \\[0.3em]
     \boldsymbol{A}^{G,q} & \boldsymbol{A}^{G,G}        
     \end{pmatrix}, \quad \boldsymbol{B} = \begin{pmatrix}
    \boldsymbol{B}^{q,q} &\boldsymbol{B}^{q,G}         \\[0.3em]
     \boldsymbol{B}^{G,q} & \boldsymbol{B}^{G,G}        
     \end{pmatrix}
\end{align}
where

\noindent\begin{minipage}{.5\linewidth}
\begin{align}
A^{q,q}_{IJ}    &=  \left<\Psi\left| \left[\hat{q}_{I},\hat{H}\left(\vec{\kappa}_{\text{opt}}\right),\hat{q}^\dagger_{J} \right] \right|\Psi\right> \nonumber \\
 A^{q,G}_{IJ}    &=  \left<\Psi\left| \left[\hat{q}_{I}, \hat{H}\left(\vec{\kappa}_{\text{opt}}\right), \hat{G}^\dagger_{J} \right] \right|\Psi\right> \nonumber  \\
  A^{G,q}_{IJ}    &= \left(A^{q,G}_{JI}\right)^*  \nonumber  \\
  A^{G,G}_{IJ}    &=  \left<\Psi\left| \left[\hat{G}_{I}, \hat{H}\left(\vec{\kappa}_{\text{opt}}\right),\hat{G}^\dagger_{J} \right] \right|\Psi\right>  \nonumber  \\ \nonumber
\end{align}
\end{minipage}%
\begin{minipage}{.5\linewidth}
\begin{align}
  B^{q,q}_{IJ}    &=  -\left<\Psi\left| \left[\hat{q}_{I}, \hat{H}\left(\vec{\kappa}_{\text{opt}}\right),\hat{q}_{J} \right] \right|\Psi\right> \label{eq:A_B_qq} \\
    B^{q,G}_{IJ}    &=  -\left<\Psi\left| \left[\hat{q}_{I}, \hat{H}\left(\vec{\kappa}_{\text{opt}}\right),\hat{G}_{J} \right] \right|\Psi\right> \label{eq:A_B_qG} \\
     B^{G,q}_{IJ}    &=   B^{q,G}_{JI}   \\
     B^{G,G}_{IJ}    &=  -\left<\Psi\left| \left[\hat{G}_{I}, \hat{H}\left(\vec{\kappa}_{\text{opt}}\right),\hat{G}_{J} \right] \right|\Psi\right>,  \label{eq:A_B_GG}  \\ \nonumber
\end{align}
\end{minipage}
and $\ket{\Psi} = \ket{\Psi(\vec{\theta}_{\text{opt}})} $. Note that $\boldsymbol{A}^{q,q} = (\boldsymbol{A}^{q,q})^\dagger$ and $\boldsymbol{A}^{G,G} = (\boldsymbol{A}^{G,G})^\dagger$ making $\boldsymbol{A}$ Hermitian, and  $\boldsymbol{B}^{q,q} = \left(\boldsymbol{B}^{q,q} \right)^\text{T}$ and $\boldsymbol{B}^{G,G} = \left(\boldsymbol{B}^{G,G} \right)^\text{T}$ making $\boldsymbol{B}$ symmetric, and $ \boldsymbol{E}^{[2]}$ is therefore Hermitian. 
The matrices $\boldsymbol{\Sigma}$ and $\boldsymbol{\Delta}$ have a similar sub-matrix structure defined as 

\noindent\begin{minipage}{.5\linewidth}
\begin{align}
\Sigma^{q,q}_{IJ}    &= \left<\Psi\left| \left[\hat{q}^\dagger_{I}, \hat{q}_{J} \right] \right|\Psi\right> \nonumber \\
\Sigma^{q,G}_{IJ}    &=  \left<\Psi\left| \left[\hat{G}^\dagger_{I}, \hat{q}_{J} \right] \right|\Psi\right> \nonumber  \\
  \Sigma^{G,q}_{IJ}    &= \left(\Sigma^{q,G}_{JI}\right)^*  \nonumber  \\
 \Sigma^{G,G}_{IJ}    &= \left<\Psi\left| \left[\hat{G}^\dagger_{I}, \hat{G}_{J} \right] \right|\Psi\right>  \nonumber  \\ \nonumber
\end{align}
\end{minipage}%
\begin{minipage}{.5\linewidth}
\begin{align}
 \Delta^{q,q}_{IJ}    &=  -\left<\Psi\left| \left[\hat{q}_{I}, \hat{q}_{J} \right] \right|\Psi\right>  \label{eq:D_S_qq} \\
     \Delta^{q,G}_{IJ}    &=  -\left<\Psi\left| \left[\hat{G}_{I}, \hat{q}_{J} \right] \right|\Psi\right> \\
      \Delta^{G,q}_{IJ}    &=    \Delta^{q,G}_{JI}   \\
     \Delta^{G,G}_{IJ}    &=  -\left<\Psi\left| \left[\hat{G}_{I}, \hat{G}_{J} \right] \right|\Psi\right>, \label{eq:D_S_GG} \\ \nonumber
\end{align}
\end{minipage}
where $\ket{\Psi} = \ket{\Psi(\vec{\theta}_{\text{opt}})} $. Note that $\boldsymbol{\Sigma}^{q,q} = (\boldsymbol{\Sigma}^{q,q})^\dagger$ and $\boldsymbol{\Sigma}^{G,G} = (\boldsymbol{\Sigma}^{G,G})^\dagger$ making $\boldsymbol{\Sigma}$ Hermitian, and $\boldsymbol{\Delta}^{q,q} = -\left(\boldsymbol{\Delta}^{q,q} \right)^\text{T}$ and $\boldsymbol{\Delta}^{G,G} = -\left(\boldsymbol{\Delta}^{G,G} \right)^\text{T}$ making $\boldsymbol{\Delta}$ antisymmetric, and $\boldsymbol{S}^{[2]}$ is therefore Hermitian. Finally, the solution vector $\vec{v}_k$ can be grouped in terms of its excitation and de-excitation part, Eq. \eqref{eq_exci_op}, as
\begin{align}
\vec{v}_k = \begin{pmatrix}
     \vec{Z}_k          \\[0.3em]
      \vec{Y}_k    
     \end{pmatrix} = \begin{pmatrix}
      \vec{\mathcal{X}}^{(k)}          \\[0.3em]
       \vec{\mathcal{A}}^{(k)}           \\[0.3em]
       \vec{\mathcal{Y}}^{(k)}         \\[0.3em]
       \vec{\mathcal{B}}^{(k)}  
     \end{pmatrix}.
\end{align}

The oo-VQE state $\ket{\Psi(\vec{\theta}_{\text{opt}})}$ should also be well-described by the same set of excitation and de-excitation operators which generate the excited states, Eq. \eqref{eq_exci_op}. Otherwise, we would not expect that the excited states can be properly described either from the set of operators in  Eq. \eqref{eq_exci_op}. For instance, we may look at the gradients with respect to the exponential ansatz  $\ket{\Phi_{k}} = \exp(\hat{\mathcal{O}}^\dagger_k - \hat{\mathcal{O}}_k)\ket{\Psi(\vec{\theta}_{\text{opt}})}$ at  $\vec{v}_k = \vec{0}$, i.e.,  $E_I^{[1]} = \braket{\Psi(\vec{\theta}_{\text{opt}})| [\hat{H},\hat{Q}_I]|\Psi(\vec{\theta}_{\text{opt}})}$ where $\hat{Q}_I\in \{\hat{G}_I, \hat{q}_I\}$, which should be zero or, at least, close to the gradient convergence criterion for the oo-VQE state $\ket{\Psi(\vec{\theta}_{\text{opt}})}$.

 We can then measure the matrix elements of $\boldsymbol{A}, \boldsymbol{B}, \boldsymbol{\Sigma}, \boldsymbol{\Delta}$ on a quantum device using $N_A$ qubits by mapping the commutators into qubit operators acting on the inactive, active, and virtual part. For example, consider the $A_{IJ}^{q,G}$ elements. These can be computed using $N_A$ qubits 
 

 \begin{align}
&\left<\Psi\left(\vec{\theta}_{\text{opt}}\right)\left| \left[\hat{q}_{I},\hat{H}\left(\vec{\kappa}_{\text{opt}}\right),\hat{G}^\dagger_{J} \right] \right|\Psi\left(\vec{\theta}_{\text{opt}}\right)\right>  \\[0.3cm] 
&= \sum_i d^{(I,J)}_i\left(\vec{\kappa}_{\text{opt}}\right) \left<\Psi\left(\vec{\theta}_{\text{opt}}\right)\left|  \hat{P}^{(i)}_I \otimes \hat{P}^{(i)}_A \otimes \hat{P}^{(i)}_V\right|\Psi\left(\vec{\theta}_{\text{opt}}\right)\right> \label{eq_AqGIJ}  \\[0.3cm] 
&= \sum_i \mathcal{D}^{(I,J)}_i\left(\vec{\kappa}_{\text{opt}}\right)
\left<A\left(\vec{\theta}_{\text{opt}}\right)\left|\hat{P}^{(i)}_{A}\right|A\left(\vec{\theta}_{\text{opt}}\right)\right>,
 \end{align}
where $\ket{\Psi(\vec{\theta}_{\text{opt}})}$ can be written as a product of the inactive, active, and virtual space, Eq. \eqref{eq:wf_tot}, the coefficients $d^{(I,J)}_i(\vec{\kappa}_{\text{opt}})$  depend on the Hamiltonian coefficients and the coefficients from the mapping of the operators from  fermionic to Pauli operators, $\mathcal{D}^{(I,J)}_i(\vec{\kappa}_{\text{opt}}) = d^{(I,J)}_i(\vec{\kappa}_{\text{opt}}) (-1)^{ \left(\# \hat{\sigma}_Z\right)^{(i)}_I}$ with $ \left(\# \hat{\sigma}_Z\right)^{(i)}_I$  the number of Pauli-\emph{Z} operators in the inactive space (\emph{I}), and $\ket{A(\vec{\theta}_{\text{opt}})}$ denotes the active space part.  Similar calculations can be done for the remaining elements of $\boldsymbol{A}, \boldsymbol{B}, \boldsymbol{\Sigma}, \boldsymbol{\Delta}$. From the matrix element evaluations, the generalized eigenvalue equation \eqref{eq:secular_eq} is constructed, and its eigenvalues are then classically solved. It is, however, not necessary to compute all the matrix elements in Eqs. \eqref{eq:A_B_qq}--\eqref{eq:A_B_GG} and \eqref{eq:D_S_qq}--\eqref{eq:D_S_GG}, since many will always evaluate to zero and many can be simplified.  For example, we find $\boldsymbol{\Delta} = \textbf{0}$ and $\Sigma^{q,G}_{IJ} = \Sigma^{G,q}_{IJ} = 0$ $\forall I,J$, which  simplifies the $\textbf{S}^{[2]}$ matrix. The elements of  $\boldsymbol{A}$ and $\boldsymbol{B}$ can  be simplified using the fact that $\hat{q}^\dagger_I\ket{\Psi(\vec{\theta}_{\text{opt}})} = 0$ and that $\hat{G}_I$ only acts in the active space, reducing the cost of evaluating the $\boldsymbol{E}^{[2]}$ matrix. For example, the $A^{q,G}_{IJ}$ elements can be obtained by evaluating the expectation values of $\hat{G}^\dagger_J \hat{H} \hat{q}_I$, $\hat{H} \hat{q}_I\hat{G}^\dagger_J $, and  $\hat{H} \hat{G}^\dagger_J \hat{q}_I $ instead of $[ \hat{q}_I,\hat{H},\hat{G}^\dagger_J]$. The working equations can be found in Appendix \ref{app:derivation}.

In this work, each element of the  matrices is explicitly constructed, and the full generalized eigenvalue equation is solved classically. 
This approach eventually becomes computationally intractable when targeting larger active spaces due to the cost associated with the explicit construction of the $\boldsymbol{A}, \boldsymbol{B}, \boldsymbol{\Sigma}, \boldsymbol{\Delta}$ matrices and the classical diagonalization. 
If only a few excitation energies are desired, matrix-free methods such as the Davidson algorithm~\cite{davidson_iterative_1975} can be used instead, requiring only the ability to carry out the linear transformations $\vec{\sigma} = \boldsymbol{E}^{[2]}\vec{s}$ and $\vec{\tau}= \boldsymbol{S}^{[2]} \vec{s}$ for an arbitrary trial vector $\vec{s}$.
Such methods are well-known and widely applied in conventional quantum chemistry~\cite{olsen_solution_1988,jorgensen_linear_1988,helmich_casscf_2019}.
The Davidson method was recently applied to extract excitation energies with the q-sc-EOM method~\cite{kim_two_2023} (Table \ref{tbl:overview}).

\subsection{The self-consistent manifold}

In recent years, the so-called self-consistent operators have been explored in the context of quantum computing and EOM (Table \ref{tbl:overview}). Mukherjee and co-workers proposed the  self-consistent manifold $\{\hat{S}_I \}$  $\cup $ $\{\hat{S}^\dagger_I \}$ which is a unitary transformation of the operator manifold of $\hat{G}_I$~\cite{prasad_aspects_1985,datta_consistent_1993}

\begin{align}
\hat{S}_I &= \hat{U} \hat{G}_I    \hat{U}^\dagger \label{eq:sc_G},
\end{align}
where $\hat{U} = \mathrm{e}^{\hat{\sigma}(\vec{\theta}_{\text{opt}})}$ is the unitary operator that prepares the ansatz in the active space. The orbital rotation operators $\{\hat{q}_I\}$ and $\{\hat{q}^\dagger_I\}$ are not transformed. 
This yields a similar generalized eigenvalue equation as presented in Eq. \eqref{eq:secular_eq} but with the substitution $\hat{G}_I \rightarrow \hat{S}_I$. The self-consistent manifold is preferable over the primitive manifold  $\{\hat{G}_I \}$  $\cup $ $\{\hat{G}^\dagger_I \}$ for several reasons. First, upon expansion of Eq.~\eqref{eq:sc_G}, high-order disconnected excitation operators can be reached, yielding better excitation energies. Second, the killer condition, $\hat{O}_k\ket{\Psi_0} = 0$ $\forall k$,  is closer to being fulfilled because the $\textbf{B}$ matrix is closer to being the zero matrix. Third, some matrix elements simplify, e.g., $A^{S,S}_{IJ} = \braket{\text{HF} | \hat{G}^\dagger_J \hat{U}^\dagger \hat{H} \hat{U} \hat{G}_I |\text{HF} } - \delta_{IJ}E_{\text{VQE}}$ compared to $A^{G,G}_{IJ}$ in Eq.~\eqref{eq:app:AGG},  which reduces the computational time. A practical computational approach was proposed in Asthana \emph{et al.}~\cite{asthana_quantum_2023} to obtain the elements $A^{S,S}_{IJ}$ following a VQE procedure without resorting to expensive algorithms such as the Hadamard test circuit~\cite{aharonov_polynomial_2009,mitarai_methodology_2019}. However, a requirement for that approach is that the operator is  Hermitian. For example, the operator $\hat{Q}^\dagger_I \hat{U}^\dagger \hat{H} \hat{U} \hat{Q}_J$ is Hermitian, and therefore the real part of its matrix elements can be written as

\begin{align}
\text{Re}\left[M_{IJ}\right] &=\frac{1}{2} \left(\left<0_{\text{ref}}\left| \left(\hat{Q}^\dagger_I + \hat{Q}^\dagger_J\right) \hat{U}^\dagger \hat{H} \hat{U} \left(\hat{Q}_I + \hat{Q}_J\right) \right|0_{\text{ref}}\right>\right.\label{eq:ReMIJ}\\\nonumber
&\quad\left.- \left<0_{\text{ref}}\left| \hat{Q}^\dagger_I \hat{U}^\dagger \hat{H} \hat{U} \hat{Q}_I \right|0_{\text{ref}}\right> - \left<0_{\text{ref}}\left| \hat{Q}^\dagger_J \hat{U}^\dagger \hat{H} \hat{U} \hat{Q}_J \right|0_{\text{ref}}\right>\right),
\end{align}
where $\hat{Q}_I\in \{ \hat{G}_I, \hat{q}_I \}$.
We see from Eq. \eqref{eq:ReMIJ} that the off-diagonal elements can be computed in terms of expectation values, which is suitable for VQE. However, in the active space approximation and using the self-consistent manifold, we would need to compute expectation values of operators such as  $\hat{H} \hat{U}\hat{G}^\dagger_{J} \hat{U}^\dagger  \hat{q}_{I}$, which is non-Hermitian. Although the Hadamard test circuit could be used to compute these elements, it is not a near-term algorithm, and on this basis we will in the current work rely on the primitive operator manifold. There are however other operator transformations than the primitive $\{\hat{G}_I \}$  $\cup $ $\{\hat{G}^\dagger_I \}$ and self-consistent $\{\hat{S}_I \}$  $\cup $ $\{\hat{S}^\dagger_I \}$  manifold. We recently explored this in Ref.~\cite{ziems_options_2023}, where we investigated eight different operator transformations for near and long-term quantum computing.


\section{Computational details}
\label{sec:com_details}

We used a noise-free simulator to validate the performance and accuracy of the oo-VQE-qEOM approach developed in this work. Because the focus of this work is benchmarking of the developed oo-VQE-qEOM method, noise-free simulations are more suitable for our purposes.

The molecular integrals for the Hamiltonian were provided by the PySCF package~\cite{sun_recent_2020}, and the mapping to the Pauli representation of the Hamiltonian was performed through the OpenFermion program~\cite{mcclean_openfermion_2020} using a Jordan-Wigner transform. 
The quantum circuits for the qEOM calculations were constructed using the Tequila software package~\cite{kottmann_tequila_2021} with qulacs~\cite{Suzuki_qulacs_2021} as simulation backend. The oo-VQE wavefunction  $\ket{\Psi(\vec{\theta}_{\text{opt}})}$ is calculated using the UCCSD ansatz and one trotter step, as implemented in Tequila using the PySCF interface~\cite{kottmann_tequila_a_platform_2020}. Full configuration interaction (FCI) and CASSCF calculations were performed using the Dalton software package~\cite{aidas_dalton_2013, dalton_2022}, and the excitation energies and the excited state properties, e.g., oscillator strengths, were extracted from the linear response (LR) module in Dalton.

The qEOM procedure begins after the oo-VQE step with the UCCSD ansatz. For the oo-VQE algorithm, we first run a UCCSD optimization in the active space, i.e., the conventional VQE algorithm, setting the starting circuit angles to zero and the starting orbitals to the MP2 natural orbitals~\cite{Jensen_second-order_1988,Mller_note_1934}. Next, we run a oo-UCCSD optimization using the optimized angles from the previous UCCSD optimization as the starting circuit angles and the same MP2 natural orbitals as the starting orbitals. The optimization of the circuit angles $\vec{\theta}$ is carried out via the Broyden-Fletcher-Goldfarb-Shannon (BFGS) implemented in SciPy package~\cite{virtanen_scipy_2020} with a gradient convergence criterion of $10^{-4}$. We refer to Appendix \ref{sec:app:oovqe} for an explanation of the connection between the error in the oo-VQE ansatz and the error in the molecular properties.

The one-photon electronic absorption spectrum~\cite{barone_computational_2012} was calculated using a Gaussian convolution of the excitation energies and oscillator strengths through the Dalton Project~\cite{olsen_dalton_2020} with a broadening factor of 0.4 eV. The electronic circular dichroism (ECD) spectrum in the length gauge was calculated using a Gaussian convolution of the excitation energies and rotational strengths with a broadening factor of 0.4 eV. 
We refer to Ref.~\cite{rizzo_response_2011} (Eqs. 2.56 and 2.57) for more information about absorption and ECD spectra. 
Computations are carried out for electronic excitation energies and oscillator strengths at the equilibrium geometries for BeH$_2$ and H$_2$O, and electronic excitation energies and rotational strengths for H$_4$. The electronic excitation energies and the excited state amplitudes are obtained by solving the generalized eigenvalue equation in Eq.~\eqref{eq:secular_eq}, where the matrix elements (working equations) can be found in Appendix \ref{app:derivation}. We show in Appendix \ref{app:os} how to compute the oscillator and rotational strengths on a quantum device. For BeH$_2$, H$_2$O, and H$_4$, the active space was chosen to be four electrons and four spatial orbitals (eight spin orbitals) for both oo-UCCSD and CASSCF,  i.e., the number of qubits is 8.
Calculations were performed using different basis sets: STO-3G~\cite{hehre_self-consistent_1969,hehre_self-consistent_1970}, 6-31G basis~\cite{binkley_self-consistent_1977,dill_self-consistent_1975,ditchfield_self-consistent_1971,hehre_self-consistent_1972}, and, 6-31G*~\cite{binkley_self-consistent_1977,dill_self-consistent_1975,ditchfield_self-consistent_1971,hehre_self-consistent_1972,gaussian09,hariharan_the_1973}.

\section{Results and discussion}

In this section, we present electronic absorption spectra for BeH$_2$ and H$_2$O and ECD spectrum for H$_4$. We compare the spectra using CASSCF with linear response (CASSCF-LR), FCI-LR, and the oo-VQE-qEOM method presented in this work using single and double excitation and de-excitation operators (qEOMSD). We also test our method for different basis set sizes. All simulations use an ideal, noise-free setting, and we do not consider the effect of sampling noise from the VQE measurements, since the focus of this work is benchmarking of the developed oo-VQE-qEOM method.

\begin{figure}
\centering  
\includegraphics[width=1.0\textwidth]{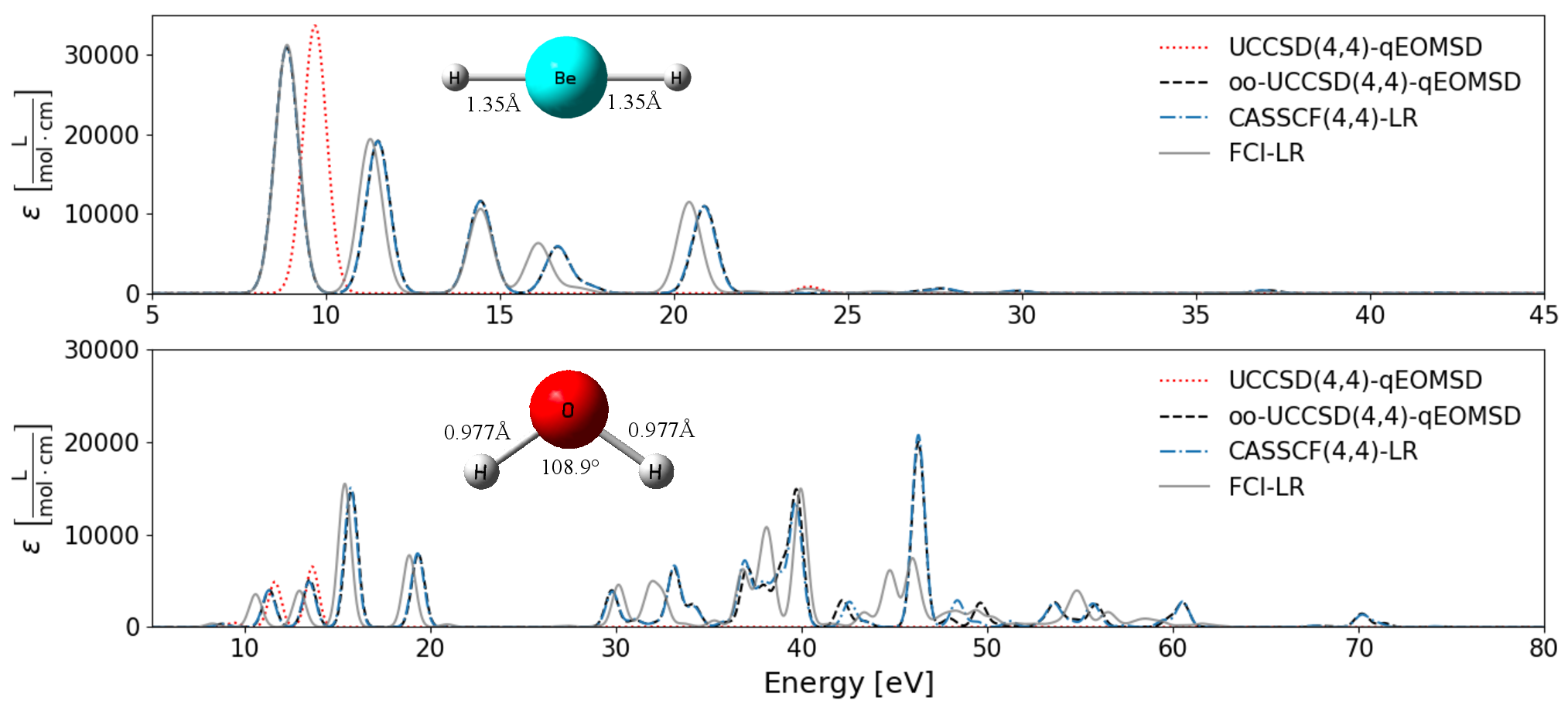}
\caption{One-photon  electronic absorption spectra for BeH$_2$ and H$_2$O in the 6-31G basis set at the equilibrium geometries, obtained from the oo-VQE-qEOM method presented in this work. We use single and double excitation and de-excitation operators (qEOMSD) and the orbital-optimized UCCSD ansatz (oo-UCCSD) to approximate the ground state.}
\label{fig:qeom}
\end{figure}

Figure \ref{fig:qeom} shows the electronic absorption spectra calculated using Gaussian convolution, with excitation energies and oscillator strengths from FCI-LR, CASSCF-LR, and oo-UCCSD-qEOMSD, for BeH$_2$ and H$_2$O. The oo-UCCSD-qEOMSD absorption spectra are almost identical to those calculated using CASSCF-LR. The oo-UCCSD-qEOMSD produces absorption spectra of the same quality as CASSCF-LR but with fewer configuration parameters in the active space (14 for qEOMSD and 19 for CASSCF-LR),\rev{ i.e., qEOMSD omits the triple and quadruple operators.} \rev{Additionally, we compare with UCCSD-qEOM, where conventional VQE is used to estimate the ground state (i.e., without orbital optimization) and the excitation operators in Eq. \eqref{eq_exci_op} are without the orbital rotation operators. The UCCSD-qEOM method therefore operates solely within the active space. We note the significance of incorporating orbital rotation operators outside and between the active space, as evidenced by the red dotted line, which considerably differs from CASSCF-LR. This is to be expected, given the relatively small active space of (4,4), and therefore the exclusion of significant information in UCCSD-qEOM.} For BeH$_2$, the oo-UCCSD-qEOMSD spectra are in very good agreement with the corresponding FCI-LR spectra, especially for the lower-lying excitations. This is to be expected since the active space includes all valence electrons. In contrast, for the H$_2$O absorption spectra, only four out of eight valence electrons are included in the active space. In this case, even the lowest-lying excitations are shifted by about 1 eV compared to the FCI spectra.

\begin{figure}
\centering  
\includegraphics[width=1.0\textwidth]{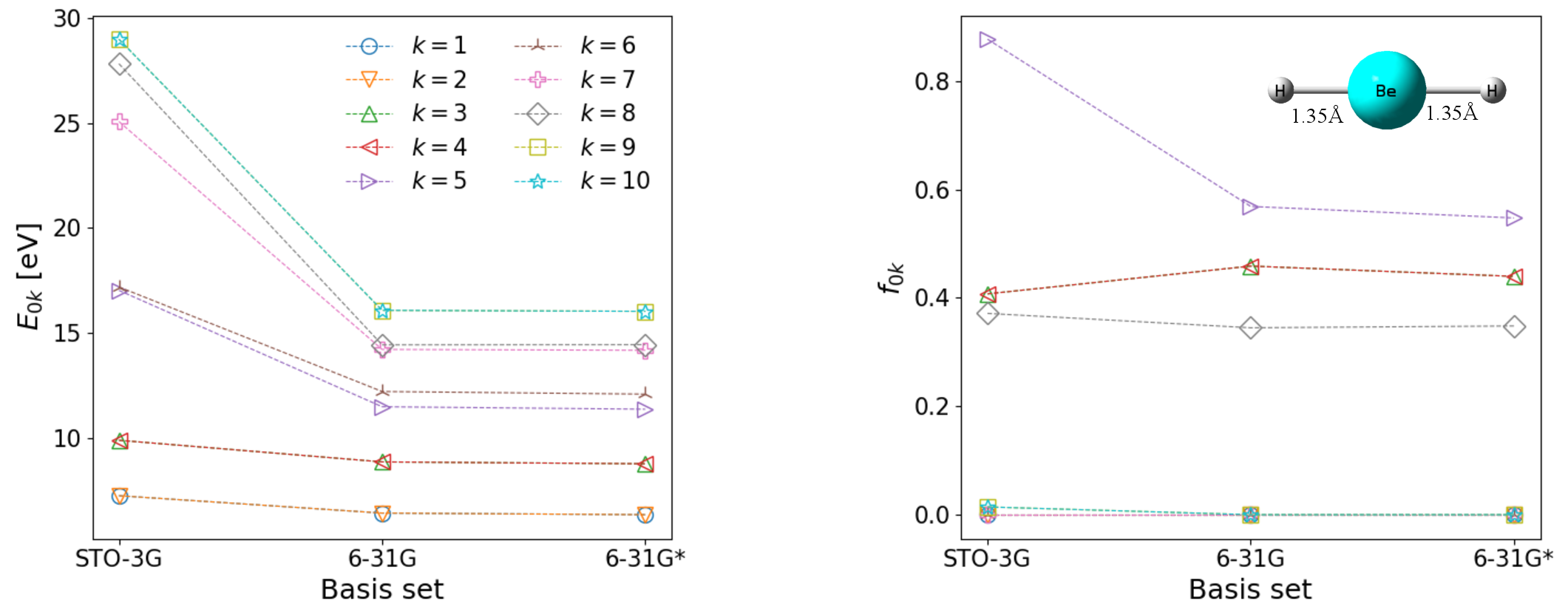}
\caption{Excitation energies (left) and oscillator strengths (right) for BeH$_2$ at the equilibrium geometry, obtained from the oo-UCCSD(4,4)-qEOMSD method.}
\label{fig:E0k_OS}
\end{figure}

Figure \ref{fig:E0k_OS} shows the excitation energies and oscillator strengths of BeH$_2$ as a function of basis set size. 
The excitation energies are strongly affected by increases in the basis set size, especially when going from the minimal STO-3G basis to the 6-31G basis. One might speculate that the large jump in the energy (more than 10 eV for some of the higher excited states) could be caused by additional low-lying states appearing when increasing the size of the basis set expansion. However, considering the oscillator strengths, it appears that this is not the case and that the character of the states is largely unchanged. For example, the $S_8$ state sees a decrease in the excitation energy from 27.8 eV to 14.4 eV, with the oscillator strength remaining almost unchanged.
The ability to treat larger basis set expansions is one of the important advantages of the oo-VQE-qEOM method since it allows for at least some degree of basis set convergence. We also note that larger basis sets would be required to confirm any convergence in the basis set size.

\begin{figure}
\centering  
\includegraphics[width=0.9\textwidth]{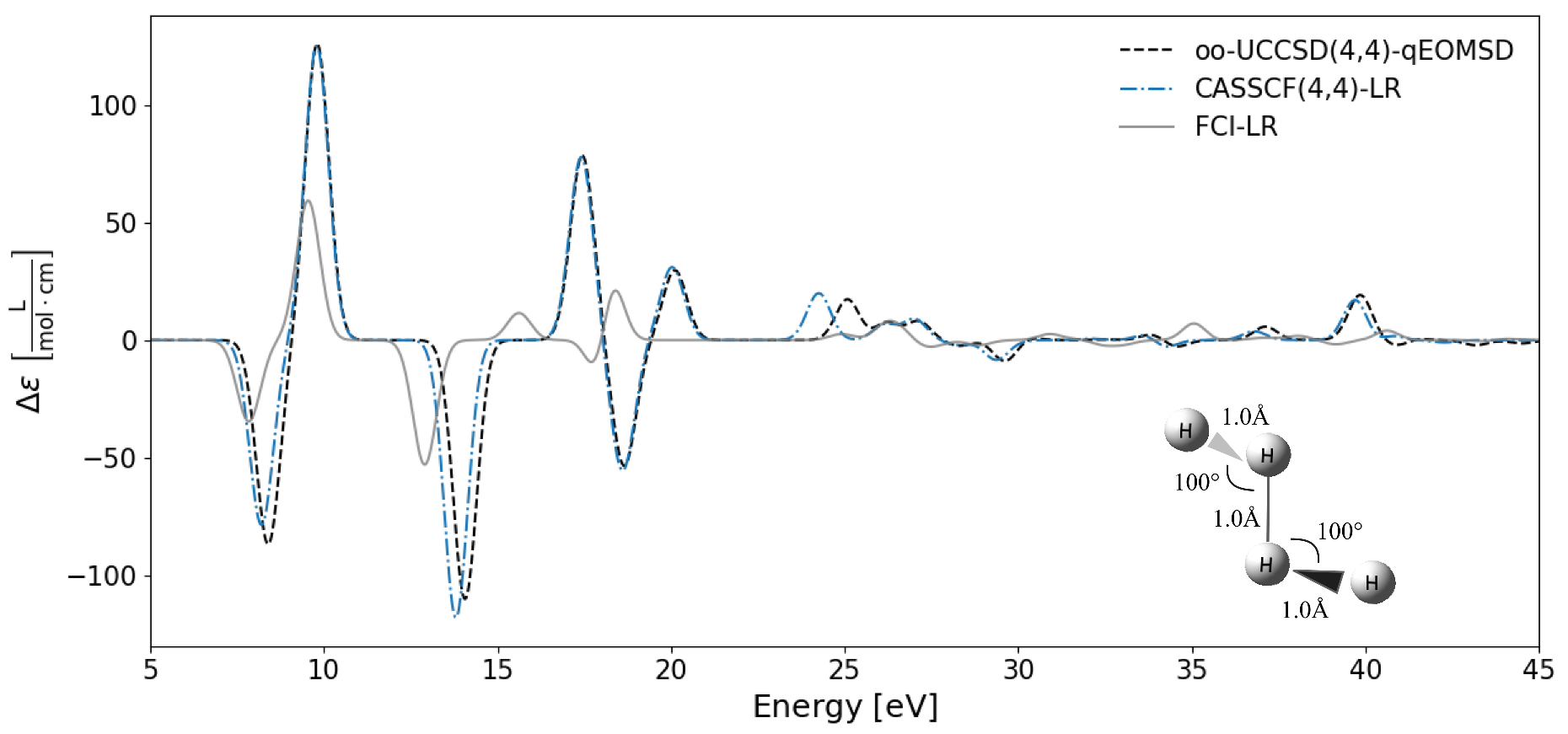}
\caption{Electronic circular dichroism  spectrum for twisted H$_4$ with dihedral angle $=120^{\circ}$, obtained from the oo-UCCSD(4,4)-qEOMSD method.}
\label{fig:ECD}
\end{figure}

\begin{figure}
\centering  
\includegraphics[width=0.9\textwidth]{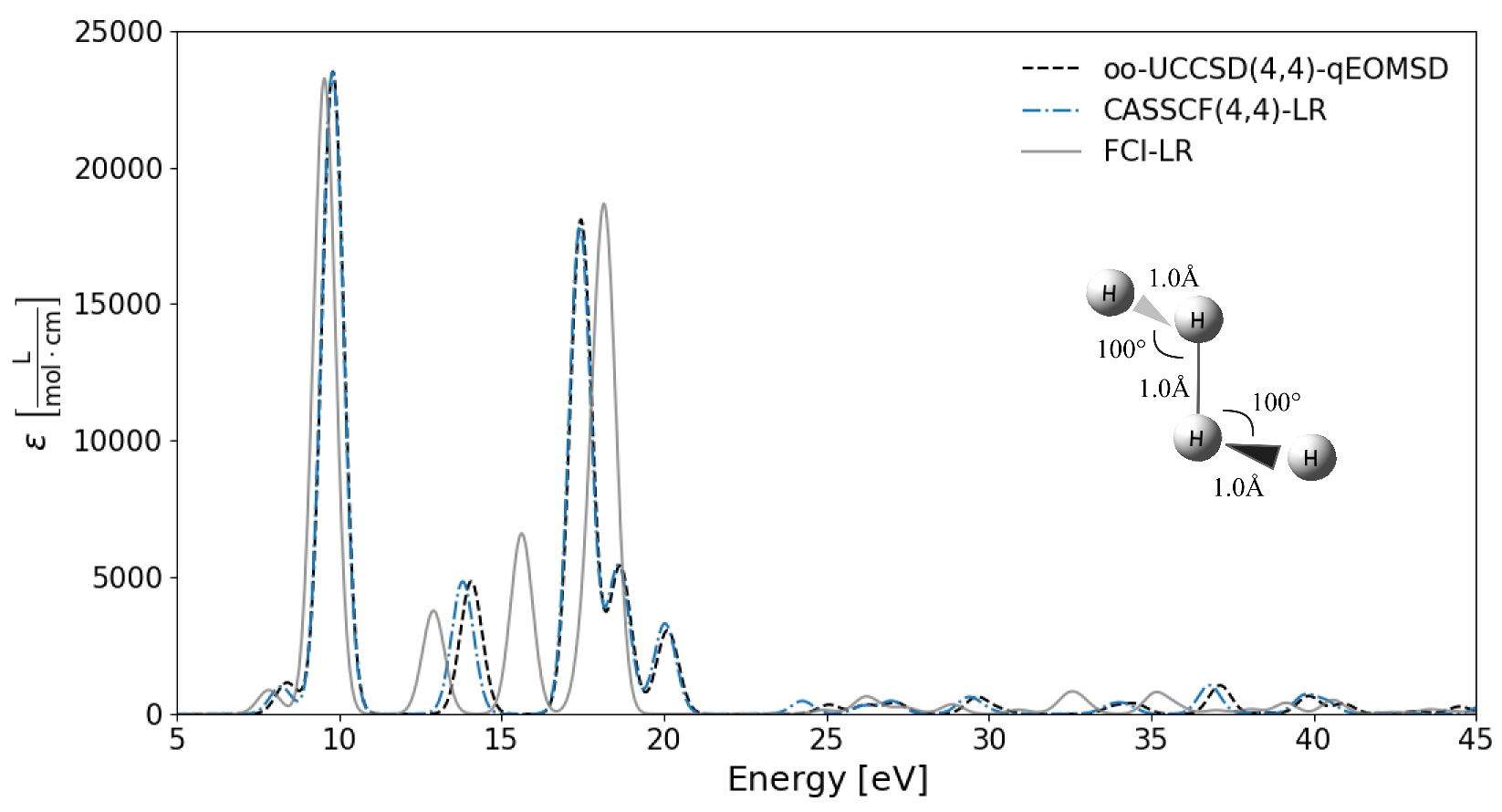}
\caption{One-photon  electronic absorption spectrum for twisted H$_4$ with dihedral angle $=120^{\circ}$, obtained from the oo-VQE-qEOM method presented in this work. We use single and double excitation and de-excitation operators (qEOMSD) and the orbital-optimized UCCSD ansatz (oo-UCCSD) to approximate the ground state.}
\label{fig:OPA_H4}
\end{figure}

To further illustrate the flexibility of our implementation, we have in addition to ordinary absorption spectra also calculated the ECD spectrum for twisted H$_4$, as shown in Fig. \ref{fig:ECD}. 
ECD originates from the differential absorption of left- and right-circularly polarized light in the frequency region of electronic transitions, and it can be regarded as the chiroptical counterpart of absorption. ECD is one of the spectroscopic techniques used to assign the absolute configuration of chiral species, since enantiomers have oppositely signed ECD spectra. The determination of ECD spectra requires the rotational strengths (Appendix \ref{app:os}). We include all four electrons in the CASSCF-LR, however, we still observe a significant disagreement with FCI.  For better agreement between CASSCF and oo-UCCSD-qEOMSD, the qEOM parameterization would require higher rank excitation operators, for this case triples (oo-UCCSDT-qEOMSDT) or even quadruples (oo-UCCSDTQ-qEOMSDTQ). In any case, oo-UCCSD(4,4)-qEOMSD completely reproduces the sign pattern of CASSCF(4,4)-LR, which is one of the critical points when simulating ECD spectra. The corresponding one-photon absorption spectrum is given in Fig. \ref{fig:OPA_H4}, which shows the same 
trend as the ECD spectrum in Fig. \ref{fig:ECD}. For example, we observe a larger disagreement around 14 eV and 25 eV both for the ECD and the one-photon absorption spectra, but otherwise is in good agreement with CASSCF.

\section{Conclusions}

In this paper, we have developed a new protocol for evaluating molecular properties for near-term quantum computers. Inspired by the quantum equation-of-motion (qEOM) algorithm, our method evaluates excitation energies and excitation amplitudes by solving a generalized eigenvalue equation. This is accomplished by an extension of the original qEOM algorithm~\cite{ollitrault_quantum_2020} by employing the orbital-optimized VQE algorithm to approximate the ground state and incorporating orbital rotations outside the active space and between the spaces in the qEOM equations to treat larger basis set expansions. The matrix elements of the generalized eigenvalue problem are then estimated directly on a quantum computer without resorting to expensive algorithms. 

We tested the method on BeH$_2$ in STO-3G, 6-31G, and 6-31G*, H$_4$ and H$_2$O in 6-31G to evaluate excitation energies, excited states, and oscillator strengths, which were then used to generate electronic absorption spectra, and for twisted H$_4$, rotational strengths and circular dichroism spectra. We demonstrate that the proposed algorithm oo-VQE-qEOM is able to reproduce CASSCF-LR calculations well using a (4,4) active space (eight spin orbitals). In future work, we will explore larger active spaces, and it is our hope that this oo-VQE-qEOM method would open the possibility of simulating larger molecular systems, perhaps biologically relevant systems, on near-term quantum computers.

\section*{Acknowledgements}
We acknowledge the financial support of the Novo Nordisk Foundation to the focused research project \textit{Hybrid Quantum Chemistry on Hybrid Quantum Computers} (HQC)$^2$ (grant number: NNFSA220080996). We thank Jakob Kottmann for helpful suggestions with respect to the Tequila~\cite{kottmann_tequila_2021} implementation of the oo-VQE-qEOM method, and Lasse Bj\o rn Kristensen for reading the manuscript and his suggestions to improve this work. 
\appendix

\section{Working equations for \textbf{A}, \textbf{B}, $\Sigma$, $\Delta$ }
\label{app:derivation}

In this section, we derive the working equations of  \textbf{A}, \textbf{B}, $\Sigma$, $\Delta$, i.e., the implemented equations,  which are simplified versions of Eqs.~\eqref{eq:A_B_qq} - \eqref{eq:A_B_GG} and Eqs.~\eqref{eq:D_S_qq} - \eqref{eq:D_S_GG}.

We start with the $\textbf{S}^{[2]}$ matrix given by 

\begin{align}
\textbf{S}^{[2]} = \begin{pmatrix}
    \boldsymbol{\Sigma} & \boldsymbol{\Delta}          \\[0.3em]
     -\boldsymbol{\Delta} ^* &  -\boldsymbol{\Sigma}^*           
     \end{pmatrix},  
\end{align}
where the matrix elements are given in  Eqs. \eqref{eq:D_S_qq} - \eqref{eq:D_S_GG}. Consider the one-electron singlet excitation operators in Eq.~\eqref{eq:q_I}. The commutator relations are given by

\begin{align}
\left[\hat{q}_I, \hat{q}_J \right]  =
    \begin{cases}
     \frac{1}{2}\left[\hat{E}_{vi}, \hat{E}_{ai}\right] = 0  & \\
      \frac{1}{2}\left[\hat{E}_{vi}, \hat{E}_{av'}\right] = -\frac{1}{2} \delta_{vv'} \hat{E}_{ai} & \\
       \frac{1}{2} \left[\hat{E}_{ai}, \hat{E}_{av}\right]= 0,  & 
    \end{cases}       \label{eq:app:qq}
\end{align}
using $\left[\hat{E}_{pq},\hat{E}_{rs}\right] = \delta_{qr} \hat{E}_{ps} - \delta_{ps} \hat{E}_{rq} $. It follows

\begin{align}
\Delta_{IJ}^{q,q}&=-\left<\Psi(\vec{\theta}_{\text{opt}}) \left|  \left[\hat{q}_{I}, \hat{q}_{J} \right]\right|\Psi(\vec{\theta}_{\text{opt}}) \right> = 0  \quad \forall I,J \label{eq:app:Dqq1}
\end{align}
since $\hat{E}^\dagger_{ai} \ket{\Psi(\vec{\theta}_{\text{opt}})} = 0$. Next, let $\hat{G}_I$ be an \emph{N}-electron excitation operator in the active space. We find

\begin{align}
\left[\hat{G}_I, \hat{G}_J \right] &= \left[\prod^I_k \hat{a}^\dagger_{a_k} \hat{a}_{i_k}, \prod^J_l \hat{a}^\dagger_{a_l} \hat{a}_{i_l}  \right] \label{eq:app:GG}\\
&=\left(\prod^I_k \hat{a}^\dagger_{a_k} \hat{a}_{i_k}\right)\left(\prod^J_l \hat{a}^\dagger_{a_l} \hat{a}_{i_l}\right) - \left(\prod^J_l \hat{a}^\dagger_{a_l} \hat{a}_{i_l}\right)\left(\prod^I_k \hat{a}^\dagger_{a_k} \hat{a}_{i_k}\right) \\
&= \left(\prod^I_k \hat{a}^\dagger_{a_k} \hat{a}_{i_k}\right)\left(\prod^J_l \hat{a}^\dagger_{a_l} \hat{a}_{i_l}\right) - \left(\prod^I_k \hat{a}^\dagger_{a_k} \hat{a}_{i_k}\right)\left(\prod^J_l \hat{a}^\dagger_{a_l} \hat{a}_{i_l}\right) \\
&= 0, 
\end{align}
because $a\neq i$, where \emph{a} indicates an unoccupied (virtual) orbital and \emph{i} indicates an occupied Hartree-Fock orbital, and moving an entire product will always move an even number of operators, therefore no sign change. Thus,

\begin{align}
\Delta_{IJ}^{G,G}&=-\left<\Psi(\vec{\theta}_{\text{opt}}) \left|  \left[\hat{G}_{I}, \hat{G}_{J} \right]\right|\Psi(\vec{\theta}_{\text{opt}}) \right> = 0  \quad \forall I,J. \label{eq:app:Sigma_GG}
\end{align}
Next, since $ \hat{q}^\dagger_I \ket{\Psi(\vec{\theta}_{\text{opt}})}  =0$ and $\hat{G}_I$ only acts in the active space, it follows

\begin{align}
\Delta_{IJ}^{q,G}&=-\left<\Psi(\vec{\theta}_{\text{opt}})\left|  \left[\hat{G}_{I},\hat{q}_{J} \right]\right|\Psi(\vec{\theta}_{\text{opt}}) \right>   = 0 \quad \forall I,J \\[0.3cm]
\Sigma_{IJ}^{q,G}&=\left<\Psi(\vec{\theta}_{\text{opt}}) \left|  \left[\hat{G}^\dagger_{I},\hat{q}_{J} \right]\right|\Psi(\vec{\theta}_{\text{opt}}) \right>   = 0 \quad \forall I,J.
\end{align}
The remaining matrix elements which we need to evaluate are

\begin{align}
\Sigma_{IJ}^{q,q}&=\left<\Psi(\vec{\theta}_{\text{opt}}) \left|  \left[\hat{q}^\dagger_{I},\hat{q}_{J} \right]\right|\Psi(\vec{\theta}_{\text{opt}}) \right>= \left<\Psi(\vec{\theta}_{\text{opt}}) \left| \hat{q}^\dagger_{I}  \hat{q}_{J} \right|\Psi(\vec{\theta}_{\text{opt}}) \right>  \\[0.3cm]    
\Sigma_{IJ}^{G,G} &=\left<\Psi(\vec{\theta}_{\text{opt}})\left| \left[\hat{G}^\dagger_{I},  \hat{G}_{J}\right] \right|\Psi(\vec{\theta}_{\text{opt}}) \right>,  
\end{align}
where it is not straightforward to simplify $\Sigma_{IJ}^{G,G}$. To summarize, the working equations for $\textbf{S}^{[2]}$ are

\begin{align}
 \boldsymbol{\Delta} = \textbf{0}, \quad     \boldsymbol{\Sigma}  =     \begin{pmatrix}
   \boldsymbol{\Sigma}^{q,q} &  \textbf{0}      \\[0.3cm]
     \textbf{0}    & \boldsymbol{\Sigma}^{G,G}  
     \end{pmatrix}.
\end{align}
For the $\textbf{E}^{[2]}$ matrix, using $\hat{q}^\dagger_I \ket{\Psi(\vec{\theta}_{\text{opt}})}  =0$, $\hat{G}_I$ only acts in the active space, and Brillouin's theorem (in Eq. \eqref{eq:app:Bqq1}), 

\begin{align}
\braket{\Psi(\vec{\theta}_{\text{opt}})|\hat{H}(\vec{\kappa}_{\text{opt}})\hat{q}_I|\Psi(\vec{\theta}_{\text{opt}})} = 0,
\end{align}
we can make the following simplifications

\begin{align}
A^{q,q}_{IJ} &= \frac{1}{2} \left(  \left<\Psi(\vec{\theta}_{\text{opt}}) \left| \left[\hat{q}_{I}, \left[\hat{H}\left(\vec{\kappa}_{\text{opt}}\right),\hat{q}^\dagger_{J} \right] \right] \right|\Psi(\vec{\theta}_{\text{opt}})\right> +  \left<\Psi(\vec{\theta}_{\text{opt}})  \left| \left[\hat{q}^\dagger_J, \left[\hat{H}\left(\vec{\kappa}_{\text{opt}}\right),\hat{q}_{I} \right] \right] \right|\Psi(\vec{\theta}_{\text{opt}})\right> \right)   \\
 &=  \frac{1}{2} \left( 2  \left<\Psi(\vec{\theta}_{\text{opt}}) \left| \hat{q}^\dagger_{J} \hat{H}\left(\vec{\kappa}_{\text{opt}}\right)\hat{q}_{I} \right|\Psi(\vec{\theta}_{\text{opt}})\right> 
 - \left<\Psi(\vec{\theta}_{\text{opt}}) \left| \hat{q}^\dagger_{J} \hat{q}_{I} \hat{H}\left(\vec{\kappa}_{\text{opt}}\right)\right|\Psi(\vec{\theta}_{\text{opt}})\right> \right. \nonumber \\  
 &\left.-\left<\Psi(\vec{\theta}_{\text{opt}})\left|  \hat{H}\left(\vec{\kappa}_{\text{opt}}\right)\hat{q}^\dagger_{J}  \hat{q}_{I}\right|\Psi(\vec{\theta}_{\text{opt}})\right>  \right)\\[0.3cm]
A^{q,G}_{IJ} &= \frac{1}{2} \left(  \left<\Psi(\vec{\theta}_{\text{opt}}) \left| \left[\hat{q}_{I}, \left[\hat{H}\left(\vec{\kappa}_{\text{opt}}\right),\hat{G}^\dagger_{J} \right] \right] \right|\Psi(\vec{\theta}_{\text{opt}})\right> +  \left<\Psi(\vec{\theta}_{\text{opt}})\left| \left[\hat{G}^\dagger_{J}, \left[\hat{H}\left(\vec{\kappa}_{\text{opt}}\right),\hat{q}_{I} \right] \right] \right|\Psi(\vec{\theta}_{\text{opt}})\right> \right)  \\
 &= \frac{1}{2} \left( 2  \left<\Psi(\vec{\theta}_{\text{opt}}) \left| \hat{G}^\dagger_{J} \hat{H}\left(\vec{\kappa}_{\text{opt}}\right)\hat{q}_{I} \right|\Psi(\vec{\theta}_{\text{opt}})\right> - \left<\Psi(\vec{\theta}_{\text{opt}})\left| \hat{H}\left(\vec{\kappa}_{\text{opt}}\right) \hat{q}_{I} \hat{G}^\dagger_{J} \right|\Psi(\vec{\theta}_{\text{opt}})\right>    \right. \nonumber \\
 &\left. -\left<\Psi(\vec{\theta}_{\text{opt}})\left|  \hat{H}\left(\vec{\kappa}_{\text{opt}}\right)\hat{G}^\dagger_{J}\hat{q}_{I}  \right|\Psi(\vec{\theta}_{\text{opt}})\right>  \right) \label{eq:app:AqG} \\[0.3cm]
A^{G,G}_{IJ} &= \frac{1}{2} \left(  \left<\Psi(\vec{\theta}_{\text{opt}}) \left| \left[\hat{G}_{I}, \left[\hat{H}\left(\vec{\kappa}_{\text{opt}}\right),\hat{G}^\dagger_{J} \right] \right] \right|\Psi(\vec{\theta}_{\text{opt}})\right> +  \left<\Psi(\vec{\theta}_{\text{opt}}) \left| \left[\hat{G}^\dagger_{J}, \left[\hat{H}\left(\vec{\kappa}_{\text{opt}}\right),\hat{G}_{I} \right] \right] \right|\Psi(\vec{\theta}_{\text{opt}})\right> \right) \\
 &=\frac{1}{2} \left( 2  \left<\Psi(\vec{\theta}_{\text{opt}}) \left| \hat{G}_I \hat{H}\left(\vec{\kappa}_{\text{opt}}\right)\hat{G}^\dagger_{J} \right|\Psi(\vec{\theta}_{\text{opt}})\right> + 2  \left<\Psi(\vec{\theta}_{\text{opt}})\left| \hat{G}^\dagger_{J} \hat{H}\left(\vec{\kappa}_{\text{opt}}\right)\hat{G}_{I} \right|\Psi(\vec{\theta}_{\text{opt}})\right>  \right. \nonumber \\
 &\left.- \left<\Psi(\vec{\theta}_{\text{opt}}) \left|  \hat{H}\left(\vec{\kappa}_{\text{opt}}\right) \hat{G}_I\hat{G}^\dagger_{J} \right|\Psi(\vec{\theta}_{\text{opt}})\right> - \left<\Psi(\vec{\theta}_{\text{opt}}) \left| \hat{G}^\dagger_{J} \hat{G}_I \hat{H}\left(\vec{\kappa}_{\text{opt}}\right)  \right|\Psi(\vec{\theta}_{\text{opt}})\right>  \right.\nonumber \\
 &\left.-  \left<\Psi(\vec{\theta}_{\text{opt}})\left|  \hat{H}\left(\vec{\kappa}_{\text{opt}}\right) \hat{G}^\dagger_{J}\hat{G}_I \right|\Psi(\vec{\theta}_{\text{opt}})\right> - \left<\Psi(\vec{\theta}_{\text{opt}})\left|\hat{G}_I  \hat{G}^\dagger_{J} \hat{H}\left(\vec{\kappa}_{\text{opt}}\right)  \right|\Psi(\vec{\theta}_{\text{opt}})\right> \right), \label{eq:app:AGG}
\end{align}
and

\begin{align}
B^{q,q}_{IJ} &= - \frac{1}{2} \left(  \left<\Psi(\vec{\theta}_{\text{opt}}) \left| \left[\hat{q}_{I}, \left[\hat{H}\left(\vec{\kappa}_{\text{opt}}\right),\hat{q}_{J} \right] \right] \right|\Psi(\vec{\theta}_{\text{opt}})\right> +  \left<\Psi(\vec{\theta}_{\text{opt}}) \left| \left[\hat{q}_J, \left[\hat{H}\left(\vec{\kappa}_{\text{opt}}\right),\hat{q}_{I} \right] \right] \right|\Psi(\vec{\theta}_{\text{opt}})\right> \right)  \label{eq:app:Bqq} \\
 &=\frac{1}{2} \left( \left<\Psi(\vec{\theta}_{\text{opt}}) \left|  \hat{H}\left(\vec{\kappa}_{\text{opt}}\right) \hat{q}_J\hat{q}_{I} \right|\Psi(\vec{\theta}_{\text{opt}}) \right> + \left<\Psi(\vec{\theta}_{\text{opt}})  \left|  \hat{H}\left(\vec{\kappa}_{\text{opt}}\right) \hat{q}_I\hat{q}_{J} \right|\Psi(\vec{\theta}_{\text{opt}}) \right>\right)  \label{eq:app:Bqq1} \\
 &=\left<\Psi(\vec{\theta}_{\text{opt}}) \left|  \hat{H}\left(\vec{\kappa}_{\text{opt}}\right) \hat{q}_I\hat{q}_{J} \right|\Psi(\vec{\theta}_{\text{opt}})\right> \label{eq:app:Bqq2} \\[0.3cm]
B^{q,G}_{IJ} &= -\frac{1}{2} \left(  \left<\Psi(\vec{\theta}_{\text{opt}}) \left| \left[\hat{q}_{I}, \left[\hat{H}\left(\vec{\kappa}_{\text{opt}}\right),\hat{G}_{J} \right] \right] \right|\Psi(\vec{\theta}_{\text{opt}})\right> +  \left<\Psi(\vec{\theta}_{\text{opt}})\left| \left[\hat{G}_J, \left[\hat{H}\left(\vec{\kappa}_{\text{opt}}\right),\hat{q}_{I} \right] \right] \right|\Psi(\vec{\theta}_{\text{opt}})\right> \right)\nonumber  \\
 &= \frac{1}{2} \left(  \left<\Psi(\vec{\theta}_{\text{opt}}) \left| \hat{H}\left(\vec{\kappa}_{\text{opt}}\right)\hat{G}_{J} \hat{q}_{I} \right|\tilde{\Psi}(\vec{\theta}_{\text{opt}})\right> - 2\left<\Psi(\vec{\theta}_{\text{opt}})\left| \hat{G}_{J} \hat{H}\left(\vec{\kappa}_{\text{opt}}\right) \hat{q}_{I} \right|\Psi(\vec{\theta}_{\text{opt}})\right>   \right. \\
 & \left. +\left<\Psi(\vec{\theta}_{\text{opt}})\left|  \hat{H}\left(\vec{\kappa}_{\text{opt}}\right)\hat{q}_{I} \hat{G}_{J} \right|\Psi(\vec{\theta}_{\text{opt}})\right>  \right) \\[0.3cm]
 B^{G,G}_{IJ} &=- \frac{1}{2} \left(  \left<\Psi(\vec{\theta}_{\text{opt}})\left| \left[\hat{G}_{I}, \left[\hat{H}\left(\vec{\kappa}_{\text{opt}}\right),\hat{G}_{J} \right] \right] \right|\Psi(\vec{\theta}_{\text{opt}})\right> \right. \\
 &\left.+  \left<\Psi(\vec{\theta}_{\text{opt}}) \left| \left[\hat{G}_J, \left[\hat{H}\left(\vec{\kappa}_{\text{opt}}\right),\hat{G}_{I} \right] \right] \right|\Psi(\vec{\theta}_{\text{opt}})\right> \right) \nonumber \\
 &=- \left<\Psi(\vec{\theta}_{\text{opt}}) \left| \left[\hat{G}_{I}, \left[\hat{H}\left(\vec{\kappa}_{\text{opt}}\right),\hat{G}_{J} \right] \right] \right|\Psi(\vec{\theta}_{\text{opt}})\right>.  \label{eq:app:BGG} 
\end{align}

\section{Error propagation}
\label{sec:app:oovqe}

The oo-VQE wavefunction  $\ket{\Psi(\vec{\theta}_{\text{opt}})}$ is calculated using the UCCSD ansatz in the active space, as implemented in Tequila using the PySCF interface~\cite{kottmann_tequila_a_platform_2020}. We set the gradient convergence criterion for the circuit angles $\vec{\theta}$ to $10^{-4}$, and the error in the excitation energies is expected to be on the same order, i.e., $10^{-4}$, assuming the cause of error only enters through the approximation of the ground state\footnote{In qEOM, there will also be error associated with the excitation operator $\hat{\mathcal{O}}^\dagger_k$ not being complete.}. To see this, we split the wavefunction into its exact and error component, $\ket{\Psi(\vec{\theta}_{\text{opt}})}=\ket{\Psi_0} + \ket{\delta \Psi}$, where $\ket{\Psi_0} $ is the exact ground state and $ \ket{\delta \Psi}$ is a small variation (error) in the wave function. The expectation value of some operator $\hat{O}$ can then be written as 
\begin{align}
\left<\Psi(\vec{\theta}_{\text{opt}})\left| \hat{O} \right |\Psi(\vec{\theta}_{\text{opt}})\right> &= \left<\Psi_0 + \delta \Psi\left| \hat{O} \right |\Psi_0 + \delta \Psi\right> \\[0.3cm]
&= \left<\Psi_0 \left| \hat{O} \right |\Psi_0\right> + \left< \delta \Psi\left| \hat{O} \right |\Psi_0\right> + \left<\Psi_0 \left| \hat{O} \right | \delta \Psi \right> + \left<\delta \Psi \left| \hat{O} \right | \delta \Psi \right>, \label{eq:error_in_wf}
\end{align}
where the error in the expectation value scales linearly with the error in the wavefunction. Note that the variation of the wave function is related to the gradient at the optimal amplitudes as   $\ket{\delta \Psi} = \gamma \nabla_{\vec{\theta}_\text{opt}} \ket{\Psi(\vec{\theta})} $ where $\gamma\in\mathbb{R}^+$ is a small step size. In the special case $\hat{O} = \hat{H}$, it follows from $\delta \braket{\Psi(\vec{\theta})| \hat{H} |\Psi(\vec{\theta})} =0$ since $\ket{\Psi(\vec{\theta}_{\text{opt}})}$ is a variational optimized wave function, and the error in the energy scales quadratic in the error of the wave function.

\section{Oscillator and rotational strengths}
\label{app:os}

The intensity of the absorption band is calculated from the dimensionless dipole oscillator strengths~\cite{sauer_molecular_2011}:

\begin{align}
 f_{0k} =  \sum_i  \frac{2}{3} E_{0k} \left|\left<\Psi_0\left| \hat{\mu}_i \right|\Psi_k\right>\right|^2,
\end{align}
where $i\in\{ x,y,z\}$, and the electric transition dipole moments can be constructed as

\begin{align}
\left<\Psi_0\left| \hat{\mu}_x \right|\Psi_k\right> = -\sum_{pq} x_{pq} \frac{\left<\Psi\left(\vec{\theta}_{\text{opt}}\right)\left| \left[  \hat{E}_{pq}, \hat{\mathcal{O}}^\dagger_k \right]\right|\Psi\left(\vec{\theta}_{\text{opt}}\right)\right>  }{\sqrt{\left<\Psi\left(\vec{\theta}_{\text{opt}}\right)\left|\left[\hat{\mathcal{O}}_k, \hat{\mathcal{O}}^\dagger_k \right]\right|\Psi\left(\vec{\theta}_{\text{opt}}\right)\right>}},\label{eq:transition_dipole}
\end{align}
with $x_{pq}$ being the dipole integrals, and $ \hat{\mathcal{O}}^\dagger_k$ being constructed from the amplitude vector $\vec{v}_k$  obtained from solving the qEOM generalized eigenvalue equation. The transition dipole moments, Eq. \eqref{eq:transition_dipole}, can be computed on a quantum device using $N_A$ qubits by mapping the commutators into qubit operators acting on the inactive, active, and virtual parts,

\begin{align}
\left<\Psi\left(\vec{\theta}_{\text{opt}}\right)\left| \left[  \hat{E}_{pq}, \hat{\mathcal{O}}^\dagger_k \right]\right|\Psi\left(\vec{\theta}_{\text{opt}}\right)\right>  &= \sum_i d^{(pq,k)}_i\left<\Psi\left(\vec{\theta}_{\text{opt}}\right)\left|  \hat{P}^{(i)}_I \otimes \hat{P}^{(i)}_A \otimes \hat{P}^{(i)}_V\right|\Psi\left(\vec{\theta}_{\text{opt}}\right)\right> \\
&= \sum_i \mathcal{D}^{(pq,k)}_i
\left<A\left(\vec{\theta}_{\text{opt}}\right)\left|\hat{P}^{(i)}_{A}\right|A\left(\vec{\theta}_{\text{opt}}\right)\right>,
\end{align}
where $d^{(pq,k)}_i$  depends on the coefficients from the mapping from  fermionic to Pauli operators, $\mathcal{D}^{(pq,k)}_i = d^{(pq,k)}_i(-1)^{ \left(\# \hat{\sigma}_Z\right)^{(i)}_I}$ with $ \left(\# \hat{\sigma}_Z\right)^{(i)}_I$ is the number of Pauli-\emph{Z} operators in the inactive space, and $\ket{A(\vec{\theta}_{\text{opt}})}$ denotes the active space part, Eq. \eqref{eq:wf_tot}.

Similarly, one can generate an ECD spectra for chiral molecules by calculating rotational strengths. The intensity of the ECD band is calculated from the rotational strength

\begin{align}
 \text{rs}_{0k} =  \sum_i \left<\Psi_0\left| \hat{\mu}_i \right|\Psi_k\right> \left<\Psi_k\left| \hat{m}_i \right|\Psi_0\right>,
\end{align}
where $i\in\{ x,y,z\}$, $\hat{m}_i$ are the magnetic transition dipole moment operators, and the magnetic transition dipole moment can be constructed as

\begin{align}
\left<\Psi_0\left| \hat{m}_i \right|\Psi_k\right> = \sum_{pq} m^{(i)}_{pq} \frac{\left<\Psi\left(\vec{\theta}_{\text{opt}}\right)\left| \left[  \hat{E}_{pq}, \hat{\mathcal{O}}^\dagger_k \right]\right|\Psi\left(\vec{\theta}_{\text{opt}}\right)\right>  }{\sqrt{\left<\Psi\left(\vec{\theta}_{\text{opt}}\right)\left|\left[\hat{\mathcal{O}}_k, \hat{\mathcal{O}}^\dagger_k \right]\right|\Psi\left(\vec{\theta}_{\text{opt}}\right)\right>}} 
\end{align}
with $m^{(i)}_{pq}$ being the magnetic transition dipole integrals. Note that the rotatory strengths were obtained in the length gauge, i.e., using the electric and magnetic transition dipole moments. The gauge origin was placed at $(0,0,0)$.


\bibliographystyle{achemso}
\bibliography{refs}

\end{document}